\documentclass[runningheads]{llncs}
\usepackage[T1]{fontenc}
\usepackage{graphicx}
\usepackage{booktabs}
\usepackage{makecell}
\usepackage{multirow}
\usepackage{siunitx}
\usepackage{enumitem}
\usepackage{orcidlink}
\hypersetup{hidelinks} 

\begin{document}

\title{Cross-Subject Predictive Validity for Learning Outcomes of Delayed Start Behavior}

\titlerunning{Cross-Subject Predictive Validity of Delayed Start Behavior}

\author{
Jordan Gutterman \orcidlink{0009-0000-9873-6590} \and
Ashish Gurung \orcidlink{0000-0001-7003-1476} \and
Lee Branstetter \orcidlink{0000-0001-7835-0527} \and
Kenneth Koedinger \orcidlink{0000-0002-5850-4768} \and
Vincent Aleven \orcidlink{0000-0002-1581-6657}
}

\authorrunning{J. Gutterman et al.}

\institute{Carnegie Mellon University, Pittsburgh, PA, USA \\ \email{\{jgutterm,agurung,branstet,kk1u,va0e\}@andrew.cmu.edu}}

\maketitle              

\begin{abstract}
Behavioral detectors provide valuable insights into learner motivation and self-regulation. Among these, delayed start, a new session-level detector, has shown great promise as a valid behavioral measure that generalizes well across systems. In this paper, we examine cross-subject predictive validity of delayed start behavior. Using iReady data from 711 grade 7 students, we find delayed starts during Math practice are predictive of standardized test performance in both Math ($\beta$=.07 SD, p=.02) and English ($\beta$=.10 SD, p=<.001). Additionally, using mixture modeling and sensitivity analyses, we use a data-driven strategy to operationalize the identification of delayed starters in practice. We identify two underlying sub-groups of interest: ``early starters'' (<5 minute average delay, 20\% of students) and ``chronic delayers'' (>13 minutes average delay, 20\% of students). Relative to students in neither sub-group, early starters experienced greater growth (Math $\beta$=.11 SD, p=.07; ELA $\beta$=.15 SD, p=.02), while chronic delayers had the opposite trends (Math $\beta$=$-$.13 SD, p=0.05; ELA $\beta$=$-$.11 SD, p=0.11). Session-level measures provide a new opportunity for content-independent detectors, adding a behavioral component to the traditional usage and progress based on student engagement with content. This work aims to bridge education research with classroom practice by developing interpretable measures that align with behavioral cues teachers already use during classwork sessions to monitor and support students.
\end{abstract}

\section{Introduction}\label{intro}

Self-regulated learning helps conceptualize and interpret learner behaviors, using frameworks that describe how students plan, execute, and control their learning. AIED research has informed the development of detectors of learner behavior in digital learning platforms (DLPs), which identify both productive and unproductive patterns of engagement. These detectors capture a range of learner behaviors, including effort \cite{shih2008_effortresponse,gurung2021gmm}, gaming the system \cite{baker2008_gaming2}, off-task behavior \cite{smallwood2015_mindwander}, and procrastination \cite{agnihotri2020procrastination,cormack2020procrast_assign}. Beyond insights into productive and unproductive engagement, these behaviors are predictive of short- and long-term academic outcomes \cite{pardos2013_affect,san2014_stem_longterm}. Building on this foundation, behavioral detectors have been integrated into teaching augmentation tools (e.g., teacher-facing dashboards and augmented reality glasses) to support real-time pedagogical decision-making.

Recent trends suggest the emerging need for identifying and intervening on behaviors related to learner self-regulation. Survey research has shown a worrying trend of a rising proportion of students with lower levels of conscientiousness and diligence than prior cohorts \cite{ft2025_conscientiousness,sutin2022_personalitycovid}. There is an opportunity to use such a detector of student behavior that is relevant to educators' experience in the classroom: teachers understand the process of telling their students, ``It's time to get started''.

Advances in session-level behavioral analytics have expanded our understanding of student behavior during individual academic work within classwork sessions. Gurung et al. \cite{eb25} found that behavioral measures of self-regulation (e.g., when to start, how long to work, and when to stop) are highly predictive of learning outcomes, including end-of-year teacher grades in math and performance on standardized tests in math. These session-level measures also exhibit strong reliability and cross-system generalizability across DLPs. When combined with more established predictors of learning outcomes such as gaming the system \cite{baker2008_gaming2} and prior performance, they found that delayed start (–0.25 standard deviation [SD]) was twice as predictive as gaming-the-system (–0.12 SD) and half as predictive as prior performance (0.47 SD) in MATHia. Beyond their predictive strength and reliability, these measures are also easily interpretable and closely aligned with teaching practice, making them well-suited for integration into educator-facing tools.

In this paper, we extend the exploration of delayed start behavior by examining its generalizability across subjects, and explore practical ways to operationalize delayed-start behavior and define an interpretable, practitioner-aligned threshold for real-time use during classwork sessions. This paper aims to answer these research questions:  

\begin{enumerate}[leftmargin=.48in]
    \item[RQ 1.] Is students' delayed start behavior during math practice predictive of learning outcomes in English?
    \item[RQ 2.] Are there sub-groups of learners that are revealed in their patterns of delayed start behavior and associated learning outcomes?
\end{enumerate}

\section{Background}\label{background}

\subsection{Self-regulated learning}\label{srl}

Self-regulated learning (SRL) provides a framework for understanding how learners monitor, manage, and regulate their learning. Prominent frameworks by Zimmerman \cite{zimmerman1990_srl_defined}, Pintrich \cite{pintrich2004_srl}, and Winne and Hadwin \cite{winnehadwin1998_srl} differ in scope and emphasis, particularly with respect to the relative emphasis on individual versus contextual factors. Pintrich's model, for instance, emphasizes individual control while acknowledging group dynamics, whereas Winne and Hadwin conceptualize SRL as a recursive cycle in which students continuously define tasks, set goals, and adjust strategies in response to internal states and external conditions. Pintrich further specifies four regulatory phases (planning, monitoring, control, and reflection) that operate across cognition, motivation/affect, behavior, and context.

Effective self-regulation is fundamental to learner success, with habits forming early in development \cite{montroy2016_earlysrl}. Education research has drawn on self-regulated learning to characterize skills and behaviors underlying students' decision-making \cite{panadero2017srl_review}, while research in psychology has emphasized executive functions, including working memory, inhibitory control, and cognitive flexibility \cite{diamond2013_ef}. During learning, these regulatory capacities shape how students organize and direct their efforts across tasks and contexts. Although many such processes are internal and difficult to observe directly, their effects often manifest in how students allocate and use available time. For instance, a motivated student with well-developed self-regulation skills will likely begin tasks promptly, maintain focus, and reflect on how to optimize effort when encountering difficult content.

\subsection{Diligence and procrastination}\label{procrastination}

Beyond SRL, productive time use in learning, particularly for task initiation, has been studied through the related constructs of diligence and procrastination. Diligence is characterized by sustained effort and perseverance over time \cite{zimmerman1990_srl_defined}. Much behavioral research has explored diligence as a component of behavioral regulation within SRL, often conceptualized as effort regulation \cite{zimmerman1990_srl_defined} or linked to conscientiousness \cite{kim2016_conscientiousness}. Galla et al. \cite{galla2014_diligencetask}, however, explored diligence as a stable pattern of effective time management and timeliness, rather than as a situational response; for example, compared to their peers, a diligent student is consistently on time. The Academic Diligence Task (ADT) \cite{galla2014_diligencetask} operationalizes diligence as a latent construct inferred from DLP log data. Their diligence measure, which quantifies how students allocate time between academic and leisure activities, predicted academic outcomes and was correlated with math interest, self-efficacy, and effort regulation.

Similarly, effective time management and conscientiousness have also been studied as procrastination. Procrastination is defined as ``voluntarily delay[s] an intended course of action despite expecting to be worse off for the delay,'' \cite{steel2007_procrast_meta}. Procrastination has a long history -- warnings of the pitfalls of delays date back to Greek \cite{hesiod_procrast} and Roman \cite{cicero_procrast} times. It has been conceptualized both as a stable individual trait \cite{steel2007_procrast_meta,ferrari1995_procrast}, and as contextual task avoidance in response to situational and motivational factors \cite{milgram1992_situational}. More recent approaches propose trait-state models that integrate self-regulation and emotional regulation \cite{pychyl2016_procrast_regulate,vanhooft2025_procrast_integrate}.

Kim and Seo's meta-analysis \cite{kim2015_procrast_meta_outc} found negative correlations between procrastination and both course grades (r=-.24) and assessments (r=-.11), with the larger effect on grades likely reflecting the incorporation of work completion. Beyond academic outcomes, Steel's meta-analysis  \cite{steel2007_procrast_meta} documented broader negative effects of procrastination on health, financial well-being, team success, and on psychological constructs including anxiety, stress, and well-being. In education, studies have linked learner delays to a range of measures of academic outcomes, including course registration \cite{borchers2024_procrast_enroll}, accessing course materials \cite{agnihotri2017_procrast_impact}, and assignment submission times \cite{cormack2020procrast_assign}.

Taken together, research on diligence and procrastination indicates that timely task initiation is a critical determinant of academic success. Diligent students who consistently begin work on time tend to outperform their peers, whereas those who procrastinate in starting work experience compounded negative effects on grades and broader well-being. These complementary lines of research suggest that timeliness in learning behaviors is not incidental but foundational to effective learning and broader academic success.

\subsection{Operationalizing and aligning educational measures}\label{operationalize}

A key challenge in operationalizing measures of learner behavior and performance is the need for interpretability, relatability, and alignment with educator expectations and classroom norms \cite{siemens2012_LAgap}. Effective teacher use of data is a field on its own \cite{gummer2016teacher_data,lee2024teacher_datause}, and suggests that aligning measures with constructs that are familiar and already used by educators is one method to help overcome these challenges. Bridging the gap between research complexities and classroom implementation requires the development of measures that are both meaningful to educators and relevant to learning. This is particularly pronounced for constructs within SRL, where efforts to implement effective interventions to support learner behavior have proven difficult \cite{dignath2021_srlimpl,lodge2018_srl_op}. For example, Dignath and Veenman \cite{dignath2021_srlimpl} reviewed 17 observational studies and found that SRL-supportive practices are infrequent and unevenly implemented, and with little evidence of improvements in learning outcomes from environments that were meant to facilitate SRL.

At the same time, the increasing volume of data generated by educational platforms has resulted in educators being deluged with data. The number of tools in use continues to rise: during the 2024-25 school year, students accessed an average of 48 unique learning tools per district, up from 42 in 2022-23, while educators accessed an average of 50 tools, up from 42 in 2022-23 \cite{edtechtop40_2025}. Given existing workloads, educators lack the time to effectively utilize the tools they already have, let alone adopt new ones. This proliferation presents the field of AIED research with a dual challenge: to prioritize measures that are easily interpretable and integrable into daily practice with minimal effort, and to develop approaches that generalize across platforms and contexts, such as different subject areas.

\section{Methods}\label{methods}

\subsection{Dataset \& class sessions}

This study utilizes student log data from iReady, an adaptive math DLP. The dataset contains log data from 711 grade 7 students from a school on the West Coast of the US in school years 2022-23 and 2023-24, spread over 26 classrooms. Students were expected to practice math using iReady during 45-minute class periods. Students worked on 18,909 problems over the time period, generating 3.6 million transaction logs.  Data were collected in compliance with IRB protocols, and parents or guardians could opt their child out of research participation.

As most DLPs do not explicitly track classwork session information, we dynamically infer it using the algorithm proposed by Gurung et al. \cite{eb25}. This algorithm mines log data to identify instances in which students from the same class are concurrently practicing math. Practice instances with more than five active students from the same class during school hours are classified as classwork sessions, instances with fewer than five students are classified as independent work, and any engagement outside school hours is classified as a homework session. By relying on transaction density rather than bell schedules, this approach avoids complications arising from administrative delays (e.g., discussing assignments or upcoming tests) or pedagogical choices to use iReady practice later in the period (e.g., providing lesson overviews or introducing concepts).

\begin{table}
\centering
\caption{Descriptive statistics of sample and class sessions}\label{tab:descriptives}
\setlength{\tabcolsep}{3pt}
\begin{tabular}{lccccc}
\toprule
\makecell{Academic \\year} & Students & Classes & \makecell{Class \\sessions} & \makecell{Mean sessions \\per class} & \makecell{Mean class session\\ length (mins)}\\
\midrule
2022--23 & 325 & 12 & 372 & 31 & 49.2\\
2023--24 & 386 & 14 & 469 & 33.5 & 46.6\\
\midrule
Total & 711 & 26 & 841 & 32.3 & 47.7\\
\end{tabular}
\end{table}

In this dataset, the algorithm identified 841 classwork sessions across 26 classes. On average, 23.7 students (SD = 6.8) were active during each session, and the session length aligned with the 45-minute bell schedule (47.7 minutes, SD = 14.4). Each class averaged 32.3 (SD = 3.6) sessions over the academic year.

\subsection{Delayed start behavior}

For each class session, we calculate a student's delay from the time elapsed between the first student in the class beginning work and the given student's first work action. Anchoring to the first students' start avoids counting contextual delay factors such as those described above as delays. Two metrics that aggregate the session-level delayed starts are computed: `delayed start minutes', the average of monthly averages of minutes delayed, and `relative delayed start',  the transformed delayed start scaled within each classroom to adjust for classroom effects (e.g. accounting for teacher pedagogical choices and administrative classroom structures).

\subsection{Analysis}\label{analysis}

\subsubsection{Subject-area independence of outcomes}\label{subjectindep}

To examine the cross-subject predictive performance of delayed start behavior, we apply the model outlined in Equation \ref{eq1}. This model controls for prior state test performance in both Math and English Language Arts (ELA) and uses students' delayed start during math practice to predict standardized test outcomes separately for Math and ELA.
\begin{equation}\label{eq1}
{Y_{postscore}} = \beta_0 + \beta_1 \cdot prescore_{ela} + \beta_2 \cdot prescore_{math} + \beta_3 \cdot sessionmeasure + \varepsilon
\end{equation}
$Y_{postscore}$ is the post-test of the math or ELA standardized assessment, ${prescore_{ela}}$ and ${prescore_{math}}$ are the prior spring assessment score for math and ELA, and $sessionmeasure$ is the session-level behavioral measure (e.g. minutes of delayed start) averaged over all months of the school year. A baseline model excluding the $\beta_3$ session-level measure is estimated as a comparison.

We utilize state standardized tests administered in the spring of 2022, 2023, and 2024. Grade 6 assessments are the prior attainment scores and grade 7 serves as the outcome. All scores are standardized for interpretability: prior math score (mean = 2468, SD = 87), final math score (mean = 2469, SD = 85), prior ELA score (mean = 2496, SD = 85) and final ELA score (mean = 2510, SD = 89).

\subsubsection{Understanding the delayed start distribution and its heterogeneity}\label{heterogeneity}

For RQ2, we start by examining the distribution of students' average delayed start behavior. To assess whether the data reflect a single population or multiple latent clusters, we employ Gaussian Mixture Modeling (GMM \cite{gurung2021gmm,park2018mm_procrast}). This identifies the number and characteristics of the distribution represented in the dataset, and informs whether there are natural cut-points or clusters that merit further, interpretable sub-grouping. To explore any heterogeneous effects of delayed start on academic outcomes, we conduct a sensitivity analysis to discover benchmark values beyond which delayed starts are most predictive of gains in both academic outcomes. With Equation \ref{eq2}, the continuous delayed start measure is substituted by $delayedStartGroup_c$, a dummy variable representing whether the student's average delayed start was less than the specified cut-point $c$ (e.g., $delayedStartGroup_c = 1[{delayedStart < c}]$). Cut-points are established at integer minutes of average delayed start minutes (metric 1 described above), where at least 10\% of students are in a category.
\begin{equation}\label{eq2}
Y_{postscore} = \beta_0 + \beta_1 \cdot prescore_{ela} + \beta_2 \cdot prescore_{math} + \beta_3 \cdot delayedStartGroup_c + \epsilon
\end{equation}
Finally, we identify interpretable sub-groups inferred from the mixture modeling and sensitivity analysis. For each subgroup $k$, we create a dummy indicator for sub-group membership; using Equation \ref{eq3}, a regression is estimated with the dummy variables for inferred sub-groups. This estimates the effect for each sub-group relative to a baseline of students who are not members of any identified sub-group.
\begin{equation}\label{eq3}
Y_{postscore} = \beta_0 +  \beta_1 \cdot prescore_{ela} + \beta_2 \cdot prescore_{math} + \sum{}\beta_{k} \cdot delayedStartGroup_k + \epsilon
\end{equation}

\section{Results}\label{results}

\subsection{Content-area independence of academic outcomes} 

Using Equation \ref{eq1}, we examine whether the session-level metrics from math practice, described in section \ref{subjectindep}, predict outcomes in Math and English Language Arts (ELA). The results of the models are presented in Table 2, where each row is the output of a separately estimated regression. For delayed start, the estimated effect was -.07 SD in math and -.10 SD in ELA. There were also effects for outcomes in other session-level metrics: total time in the DLP (Math = +.11 SD, ELA = +.09 SD) and relative delayed start accounting for classroom-level effects (Math = -.13, ELA = -.15).

The greatest improvement in model fit above the baseline model for relative delayed start was in ELA outcomes, where $R^2$ increased by .02 and BIC declined by 21; there were also improvements for the delayed start minutes metric ($R^2$: +.01, BIC: -7). In math outcomes, the relative delayed start measure provided gains ($R^2$: +.01, BIC: -14) and model fit change was negligible for the delayed start minutes metric. Across the board, the ELA outcome had greater overall effect size and fit improvements for delayed start behavior metrics.

\begin{table}[h]
\centering
\caption{Estimates for models of average values of session-level metrics in iReady on Math and ELA outcomes}
\label{tab:models}
\resizebox{\linewidth}{!}{ 
\begin{tabular}{l|SSSS|SSSS}
\toprule
\multirow{2}{*}{\textbf{Metric}} & \multicolumn{4}{c|}{\textbf{Math}} & \multicolumn{4}{c}{\textbf{ELA}} \\
                & {$\beta$} & {p-value} & {$R^2$} & {BIC} & {$\beta$} & {p-value} & {$R^2$} & {BIC} \\
\midrule
Baseline & {--} & {--} & .523 & 1512.5 & {--} & {--} & .509 & 1531.7 \\
Delayed start (mins) & -.07 & .02 & .527 & 1513.4 & -.10 & <.001 & .518 & 1524.9 \\
Relative delayed start & -.13 & <.001 & .536 & 1498.9 & -.15 & <.001 & .528 & 1510.6 \\
Total time on task & .11 & <.001 & .534 & 1502.8 & .09 & .001 & .517 & 1526.6 \\
\bottomrule
\end{tabular}}
\end{table}

\subsection{Response time decomposition}

\begin{figure}
    \centering
    \includegraphics[width=\linewidth]{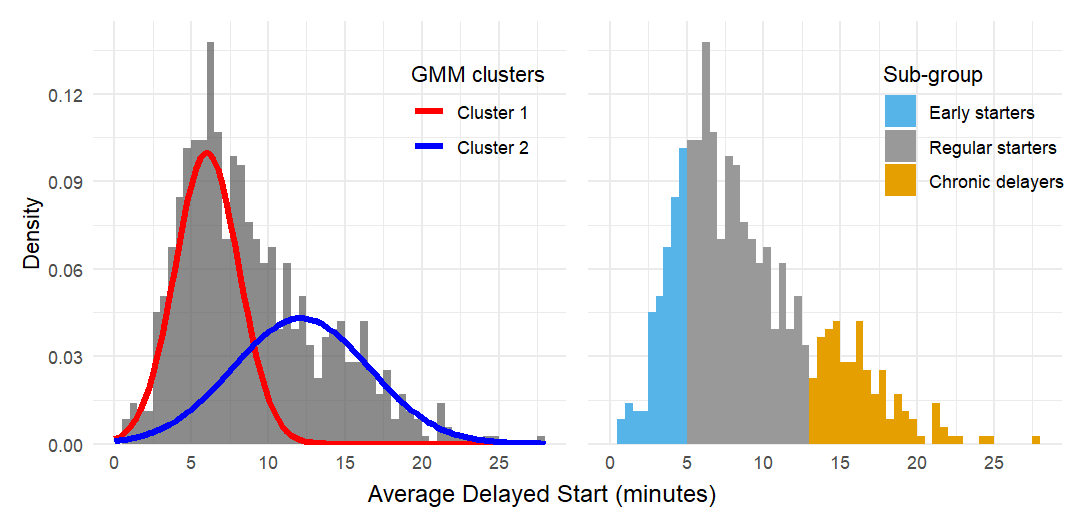}
    \caption{Histogram of student average delayed start behavior, overlaid with the two clusters found by Gaussian Mixture modeling (left); and sub-group identification as inferred by a combination of mixture modeling and sensitivity analysis (right)}
    \label{fig:gmm_combined}
\end{figure}

The distribution of students' average delayed start is illustrated in Figure \ref{fig:gmm_combined}. The mean delay was 8.96 minutes, with a long tail that extends to 27.6 minutes. The distribution peaked at approximately 6 minutes. As outlined in section \ref{analysis}, we utilize Gaussian Mixture Modeling (GMM) to determine the number of latent clusters represented in this dataset. A two-cluster model (BIC = 4076.1) performs best, outperforming both three-cluster (BIC = 4090.0), and unimodal (BIC = 4196.6) models.

The first of the two clusters represents the majority of students (n = 426, 60\%), and is centered at 6.0 minutes with an SD of 2.1 minutes. The second cluster (n = 285, 40\%) has a mean of 12.1 minutes, with a larger SD of 4.4 minutes. The point where the clusters intersect is at an average delayed start of 9.0 minutes, which is the point where a student is equally likely to belong to either cluster. On the left of Figure \ref{fig:gmm_combined}, these two suggested clusters are overlaid on the histogram of all students' average delayed start.

\vspace{-4mm} 
\vspace{-4mm}

\subsection{Sensitivity analysis}

\begin{figure}
    \centering
    \includegraphics[width=1.0\linewidth]{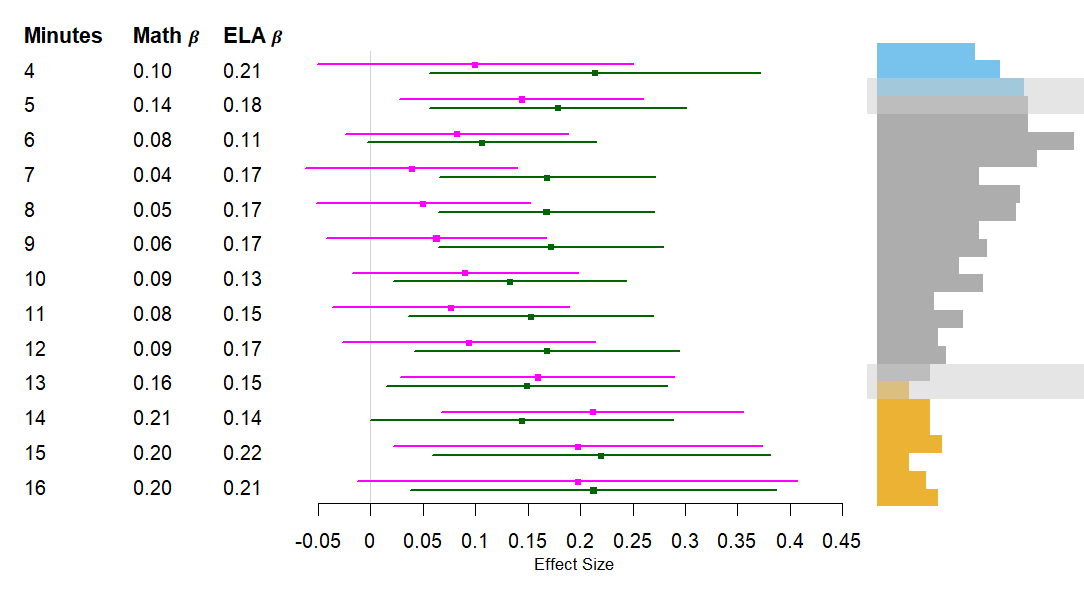} 
    \caption{Sensitivity analysis of Math outcomes (in blue) and ELA outcomes (in green) for students above cut-points of average delayed start versus those below the cut-points. On the right, the histogram of delayed start distribution is colored with suggested sub-groups, with grey bands indicating the cut-points used to define sub-group boundaries where outcomes in both subject areas are significant.}
    \label{fig:forest-sensitivity}
\end{figure}

We conduct a sensitivity analysis to examine the relationship of varying levels of delayed starts on learning outcomes using Equation \ref{eq2}. Based on the observed density of average delayed start behavior in the sample, cut-points were established at intervals of 1 minute from 4 to 16 minutes. For each cut-point, a dummy variable represented whether the student's average delayed start was below that threshold. The results are displayed in Figure \ref{fig:forest-sensitivity}.

In math outcomes, students who had 5 or fewer minutes of average delayed start had better outcomes ($\beta$ = .14, p = .016), and those with greater than 13 or 15 minutes or more had the worst outcomes. In ELA, significant positive gains were found across almost all cut-points. The cut-points with statistical significance in both subject areas were average delayed starts of less than 5 minutes (Math $\beta$ = .14, p = .02; ELA $\beta$ = .18, p = $<$.01), and at less than 13 or 15 minutes (Math $\beta$ = .16, .20, p = .02, .03; ELA $\beta$ = .15, .22, p = .03, $<$.01).\footnote{Multiple hypothesis testing correction was applied using the Benjamini-Hochberg method. All results that were statistically significant at p $<$ .05 were still significant after the corrections.} There is less variation across cut-points in ELA than in math, with values from 7 to 14 minutes falling within a range of effect sizes of .13-.17 SD.

Finally, we identified two sub-groups of interest by combining the results from the mixture modeling and sensitivity analysis. We infer two thresholds in the delayed start behavior of students: 5 minutes and 13 minutes. Each lies roughly 1 minute towards the tail from the centers of the clusters found by the mixture modeling, and coincides with the regions where the sensitivity analysis found significant gains in both subject area outcomes, with the 13 minute point also being the first found as we head towards the high delayed start tail. These define our two sub-groups of interest: ``early starters'' who began on average less than 5 minutes after the first student in the class session, and ``chronic delayers'' who began more than 13 minutes after that point. Each sub-group includes roughly 20\% of the sample.

We utilize membership in each group as dummy variables in a pair of regressions for each subject area outcome using Equation \ref{eq3}, with results summarized in Table 3. Early starters outperformed the baseline group of regular starters by .11 SD in math and .15 SD in ELA; the chronic delayers underperformed regular starters by .13 SD in math and .11 SD in ELA.

\begin{table}[h] 
\centering
\caption{Regressions on each outcome, with dummy variables for membership in sub-groups of interest}
\label{tab:final_mod}
\begin{tabular}{lc|S S|S S}
\toprule
\multirow{2}{*}{\textbf{Sub-group}} & \multirow{2}{*}{\textbf{\makecell{Proportion \\of sample}}} & \multicolumn{2}{c}{\textbf{Math}} & \multicolumn{2}{c}{\textbf{ELA}} \\
                &  & {$\beta$} & {p-value} & {$\beta$} & {p-value} \\
\midrule
Early starters ($<$ 5 mins.): & 19.7\% & .11 & {.07} & .15 & {.02} \\
Chronic delayers ($>$ 13 mins.): & 19.4\% & -.13 & {.05} & -.11 & {.11} \\
\bottomrule
\end{tabular}
\end{table}

\section{Discussion}\label{discussion}

\subsection{Cross-subject predictive validity of delayed start behavior}\label{discuss-validity}

Our examination of cross-subject predictive performance reveals that delayed starts during math practice also predicts learning outcomes in ELA ($\beta$ = .10, p = $<$.001). This suggests that delayed start behavior captures aspects of learner behavior that extend beyond subject-specific motivations or interests, and instead reflects domain-general behaviors associated with procrastination, diligence, and SRL.

The two approaches we evaluated -- a simpler raw minutes or a relativized metric -- map on to different views in the debate in the procrastination literature discussed in section \ref{procrastination}: whether the construct is a state, trait, or combination of both. The relativized measure aligns with the contextual view \cite{milgram1992_situational}, as it reflects situational comparisons within each class session; it has as support its stronger predictive power and goodness of fit. The simpler metric is associated with the trait-based approach \cite{ferrari1995_procrast}, supported by its generalizability and cross-subject validity. Together, our findings provide evidence supporting the state-trait approach where both factors contribute to the underlying construct of delayed starts.

When contextualizing our findings with prior evidence of temporal generalizability \cite{eb25}, the measure's cross-subject predictive validity further reinforces the value of delayed start behavior as a measure of self regulation. The two measures we used offer a trade-off. While the relative measure offers stronger predictive performance (Table \ref{tab:models}), the raw minutes-delayed measure is more interpretable and actionable for practitioners because it is the form the delay takes in the classroom: how many minutes students delay.

\subsection{Operationalizing the detection of delayed starters}\label{discuss-op}

Utilizing a combination of mixture modeling and sensitivity analysis demonstrates a data-driven approach to learning the most impactful sub-groups of learners' delayed start behavior. By incorporating multiple sources -- the distribution itself and multiple outcomes measures -- helps minimize overfitting of our sub-groups to any one metric. The precise sub-group definitions we found are not prescriptive thresholds that will apply to every context, as we observe from the utility of context and classroom factors in the predictive strength of the class-relativized metrics. However, the identification of heterogeneous effects in academic outcomes at each end of the distribution is a strong indication of the students most affected by this behavior, driving decisions for who an operationalization of a delayed start detector would be most effective for.

The two components found to make up the distribution aligns with findings from Park et al \cite{park2018mm_procrast}, who also used mixture modeling to classify student procrastination behaviors. They similarly found two components to best represent the procrastination behavior in assignment completion (with 36-37\% of students in the procrastination component) and a correlation between component membership and course grades.

The models including sub-group membership (reported in Table \ref{tab:final_mod}) use ``regular delayers'' as the baseline, setting a higher bar given the observed heterogeneity. While not all results are significant, it is a better approximation of the real-world potential of trying to move students between sub-groups: from chronic to regular delayers, or from regular delayers to early starters. In utilizing simpler raw minutes metrics to define them, we propose a metric that is aligned to how educators see learner delays in their classroom -- without introducing indexes or complex models, an easily understandable detector doesn't require teaching how to utilize a new index or framework. The resulting sub-groups preserve predictive value while remaining interpretable, providing a bridge between statistical models and implementable tools, and enabling the measure to be operationalized using simple benchmarks aligned with what educators observe in their classrooms.

\subsection{Implications and opportunities}\label{discuss-implications}

Our findings suggest that session-level detectors such as delayed start have promise as tools to target interventions in the classroom, aligning with educator expectations of SRL and procrastination behavior they already consider in everyday classroom management. Identified early starters are likely more motivated and could be candidates to be student mentors in small-group settings; chronic delayers could be nudged by the instructor via goal-setting around time usage, prioritization for tutoring, or focused teacher attention at the beginning of the class period.

Session-level measures like delayed start have a particular opportunity in the K-12 space. Unlike in higher education where student work is often independent, in K-12 the daily class period functions as a predetermined and bounded space for work completion. The beginning and end times of this space allows for analytically leveraging these meaningful behavioral anchors for defining measures of learner self-regulation.

Effective self-regulation and procrastination have lifelong impacts. While in this work we focused on standardized test outcomes, the ability to start and complete tasks in a responsible manner is an essential skill that goes beyond academic achievement in determining long-term outcomes such as college graduation, earnings, and life satisfaction \cite{jackson2018_teacher,moffitt2011_selfcontrol,heckman2012_softskills}. Learning how to self-regulate and develop executive functions are key skills students gain from their time in school, and this work provides opportunities to support students with behaviors that are key to both their immediate academic outcomes and their lifelong outcomes.

\section{Limitations \& future work}

This study's results were derived from one dataset that exists in a particular context: a middle school with 45 minute class periods using a particular math learning platform. In other contexts, the thresholds for detecting delayed start sub-groups we found could differ depending on changes in these and other contextual factors. For example, we would expect the benchmarks that define sub-groups to shift in a school with 60 minute class sessions.

To make results interpretable and aligned with classroom practice, we chose to use the raw delayed start in minutes metric to develop. As discussed in section \ref{discuss-op}, this leaves the predictive power of including teacher and classroom effects, contextual factors, and pedagogical choices in individual classrooms on the table. Finally, while this work provides evidence that delayed starts is a behavioral measure, it does not identify more precisely which construct it best represents.

To strengthen confidence in the use of delayed start behavior, there are several possible further validity studies that could explore the implications of interventions. While we looked at the cross-subject predictive validity of delays during math practice on ELA outcomes, future work should examine if the same is true in reverse, i.e., whether delayed starts during ELA practice predict math outcomes. Theoretically, the behavior we are detecting should have an effect on other outcomes; it would be useful to discover if delayed starts are correlated with absenteeism, tardiness, and other behavioral data. Evidence for convergent validity could come from analyzing whether and how this measure aligns with other SRL and procrastination measures, for which there are a wealth of validated survey-based measures capturing SRL constructs. This would be akin to the convergent validity with Big Five conscientiousness found by similar tools such as the Academic Diligence Task \cite{galla2014_diligencetask}.

\section{Conclusion}\label{conclusion}

In this study, we found cross-subject predictive validity of delayed start behavior from Math to English, generalizing across subjects areas. We also demonstrate data driven strategies to identify early starter and chronic delayer sub-groups that are interpretable, align with practitioner expectations and can seamlessly be integrated in classroom practice. Simple membership of the sub-groups was an indicator of academic performance in both Math and English subject areas.

More practically, our findings highlight the potential of behavioral detectors to provide interpretable metrics that align with educators' everyday observations of student behavior. In today's classrooms, session-level metrics make it possible to build up from small differences of just a few minutes in when students get started into a reliable and meaningful detector. Because this approach leverages data from digital learning platforms already in use in the classroom, it requires no new diagnostics, frames results in terms aligned with educators' experience (minutes of student delay), generalizes across subjects to impactful student behaviors, and can integrate into classroom practice.

\begin{credits}
\subsubsection{\ackname}
The authors would like to thank Curriculum Associates LLC for their partnership and collaborative efforts in this research. We especially thank Erin Banjanovic and Logan Rome of Curriculum Associates for their invaluable support and collaboration. This work was supported by the Learning Engineering Virtual Institute. The opinions, findings, and conclusions expressed in this material are those of the authors and do not necessarily reflect the views of the Institute or Curriculum Associates.
\end{credits}

\bibliographystyle{splncs04}
\bibliography{Delayed_Start}

\end{document}